\documentclass[11pt]{article}
\usepackage{graphicx}
\usepackage{amssymb}
\usepackage{amscd}
\usepackage{mathrsfs}
\usepackage{longtable,lscape}
\usepackage{amsthm}
\usepackage{color}
\usepackage{amsfonts}
\usepackage{amsmath}
\usepackage{bbm}
\usepackage{float}
\usepackage{url}

\newcommand{\Rs}{{\cal R}}
\newcommand{\Ms}{{\cal M}}

\newcommand{\Os}{{\cal O}}

\newcommand{\js}{{\vee_{}}}
\newcommand{\ms}{{\wedge_{}}}

\newcommand{\captionfonts}{\footnotesize}
\makeatletter  
\long\def\@makecaption#1#2{%
  \vskip\abovecaptionskip
  \sbox\@tempboxa{{\captionfonts #1: #2}}%
  \ifdim \wd\@tempboxa >\hsize
    {\captionfonts #1: #2\par}
  \else
    \hbox to\hsize{\hfil\box\@tempboxa\hfil}%
  \fi
  \vskip\belowcaptionskip}
\makeatother 
\begin{document}
\title{\bf The Complexity-Stability Debate, Chemical Organization Theory, and the Identification of Non-Classical Structures in Ecology} 
\author{Tomas Veloz$^{1,2,3}$\vspace{0.5 cm} \\ 
\normalsize\itshape
$^1$ Center Leo Apostel for Interdisciplinary Studies, Vrije Universiteit Brussel (Belgium)\\
\normalsize\itshape 
$^2$ Fundacion para el Desarrollo Interdisciplinario de la Ciencia, la Tecnolog\'ia y las Artes, Santiago (Chile)\\
\normalsize\itshape 
$^3$ Universidad Andres Bello, Departamento Ciencias Biol\'ogicas\\ 
\normalsize\itshape
Facultad Ciencias de la Vida, 8370146 Santiago (Chile) 
\normalsize
E-Mail: \url{tveloz@gmail.com}
}
\date{}
\maketitle
\begin{abstract}
We present a novel approach to represent ecological systems using reaction networks, and show how a particular framework called Chemical Organization Theory (COT) sheds new light on the longstanding complexity-stability debate. Namely, COT provides a novel conceptual landscape plenty of analytic tools to explore the interplay between structure and stability of ecological systems. Given a large set of species and their interactions, COT identifies, in a computationally feasible way, each and every sub-collection of species that is closed and self-maintaining. These sub-collections, called  organizations, correspond to the groups of species that can survive together (co-exist) in the long-term. Thus, the set of organizations contains all the stable regimes that can possibly happen in the dynamics of the ecological system. From here, we propose to conceive the notion of stability from the properties of the organizations, and thus apply the vast knowledge on the stability of reaction networks to the Complexity-Stability debate. As an example of the potential of COT to introduce new mathematical tools, we show that the set of organizations can be equipped with suitable joint and meet operators, and that for certain ecological systems the organizational structure is a non-boolean lattice, providing in this way an unexpected connection between logico-algebraic structures, popular in the foundations of quantum theory, and ecology. 
\end{abstract}

\section{Introduction}
The decline of the Earth’s biodiversity is a threat to the ecosystems in the planet. Ecological systems are faced with species extinctions and invasions and one fundamental question is how systems vary when they suffer these changes~\cite{Finke2004}. In particular, a major problem in theoretical ecology is to resolve how ecosystem features such as resilience, resistance, robustness, or in wider terms, stability respond to changes in species diversity, richness, connectivity, or in wider terms, complexity. 
From an abstract perspective, an ecosystem consists of a large and diverse group of species interacting in a common space in different ways. The dynamics of these interactions describe the evolution, stability, and resilience of the ecosystem~\cite{Pimm1984}. The fathers of ecology regarded as obvious the fact that more {\it entangled} ecosystems would be more stable. However, early mathematical models proven that diversity and stability can be anticorrelated for large networks~\cite{May1972}. From here, a myriad of studies have supported the two opposite views. This controversy became known as the Complexity-Stability problem, or the Complexity-Stability debate~\cite{May1973}. 

Recently, some of the most prominent figures around the Complexity-Stability debate have concluded that radically novel approaches are required to understand how different types of ecological interactions develop {\it multi-dimensional} architectures that lead to stable ecosystems. For example, in~\cite{Donohue2016} they claim:
\begin{center}
{\it We assess the scientific and policy literature and show that this disconnect is one consequence of an inconsistent and one-dimensional approach that ecologists have taken to both disturbances and stability. This has led to confused communication of the nature of stability and the level of our insight into it. Disturbances and stability are multidimensional. Our understanding of them is not.}
\end{center}

Reaction networks are the paradigmatic language of biochemical modeling. A reaction network consists of a collection of components whose interactions are determined by  consumption and production rules among the components~\cite{Lacroix2008}. Recently, the language of reaction networks has been applied beyond biochemstry. When viewed as an abstract language, reaction networks represent systems whose basic interactions consist of ` the consumption of a collection of entities producing a partially or totally new collection of entities as a result'. Thus the dynamics can be seen as `collective transformations'. Therefore, is we assume these entities of being of a not-biochemical nature, the scope of application of reaction networks is immense. Indeed, reaction networks have been applied to model the exchange of economic goods~\cite{Dittrich2005}, the influence of political decisions~\cite{Dittrich2008}, the evolution of cooperation~\cite{Veloz2014}, other game-theoretical situations~\cite{Velegol2018}, and have been recently proposed as a modeling framework for situations of multidisciplinary nature in environmental sciences~\cite{Veloz2013}, and systems theory~\cite{Veloz2017a}. Hence, we propose that reaction networks is an interesting paradigm to represent ecological interactions and ecosystems, and that can be a potential solution to the requests made by the ecological community concerning the complexity-stability debate. Namely, we aim at 
\begin{itemize}
\item Characterizing the features of current modeling languages applied to the Complexity-Stability debate to understand why these languages have not been successful in providing conclusive results.
\item Introducing the language of reaction networks for modeling ecological systems, and a particular framework called Chemical Organization Theory (COT)~\cite{Dittrich2007} to study the Complexity-Stability problem
\item Showing an example of the potential of COT to study the Complexity-Stability problem with novel mathematical tools.
\end{itemize} 

The article is organized as follows: In section~\ref{MEco} we overview the mathematical approaches most commonly used to model ecological systems, and identify their strengths and weaknesses in relation to the Complexity-Stablity debate. In section~\ref{COTM} we introduce the reaction network formalism, COT, and how it can be applied to model
ecosystems. In section~\ref{COT-CS} we discuss the potential of COT to study the Complexity-Stability debate in a way that not only overcomes the difficulties encountered with other modeling languages, but also enriches the debate by introducing novel mathematical tools. In section~\ref{QCOT} we show that the structure of stable states of an ecological system modeled using COT is in some cases a non-distributive lattice, establishing thus an unexpected link between the Complexity-Stability debate and quantum theory. 

\section{Modeling Ecosystems}
\label{MEco}
There are three fundamental representational languages for the mathematical modeling of ecosystems: Dynamical systems, Networks, and Agent based models.\\

Dynamical systems~\cite{Strogatz1999} provide a suitable framework to accurately model the interactions of a group of species. Indeed, one is able to represent the interactions at a mechanistic level, i.e. considering the specific manner in which the species interact (e.g. Lotka-Volterra systems~\cite{Chen2003}). In regards to the Complexity-Stability debate this language is interesting as it is possible to compute the dynamical evolution, and it is possible to apply the theory of dynamical systems, rich in analytic tools, to link the structure and stability of the system~\cite{May1973}. However, even moderately small dynamical systems generate extremely complicated equations that are virtually impossible to solve analytically, and very expensive to simulate computationally, and asymptotic methods are hard to analyze due to the large number of parameters involved. Thus, despite the elegance and precision of this framework, it is often inadequate to study complex ecosystems that involve large groups of diverse species. \\  

An alternative approach is to represent interactions between species as links in a network. For example, two species can be connected by a link if one species preys on the other (these networks are known as food-webs~\cite{Pimm1982}). In this way, an ecological system is represented by a network of ecological interactions~\cite{Montoya2006}. Research on the relationship between the architecture of the network and the community stability has shown that, whereas high connectance and nestedness promote stability and increases species richness in communities made up exclusively of mutualistic interactions, the stability of trophic networks is higher in modular and weakly connected architectures~\cite{Dunne2002,Kondoh2003,Thebault2000}. Therefore, there seems to be that the structure that promotes stability in an ecological network depends strongly on the type of interaction that is being considered. For a comprehensive and updated review on technical results that relate complexity and stability for the most studied types of interaction (depredation, mutualism and competition) see~\cite{Landi2018}.\\
  
Studies applying the network framework to ecology have improved our knowledge of the interplay between complexity and stability. However, networks cannot address the fact that natural communities are composed of different interaction types that operate simultaneously~\cite{Fontaine2011}. Empirical work has started to address methodologies that can incorporate different interaction types into a broader ecological network context~\cite{Melian2009,Olff2009}. These empirical studies, and recent theoretical analysis~\cite{Garcia-Callejas2018,Lurgi2015}, have opened up a new theoretical challenge in complexity-stability research. For example, since networks represent interactions as valued links, either positive or negative depending on how interactions affects the species linked, it is not clear how to value different kinds of positive or negative interactions. Another important problem is that networks can not provide a mechanistic description of how species interact, and consider only two-species interaction, while in some ecological interactions other species not considered in the link can play a contextual role.
 
There are some attempts to improve the network modeling to overcome these difficulties, but it seems to be a very hard problem. For example, in~\cite{Pilosof2017} they present a generalized version of the network-based modeling, called multi-layered framework, where different types of links represent different types of interactions, so they encompass multiple ecological interactions. In fact, this is the most advanced network-based theoretical framework available in the literature to our understanding. However, they accept solid drawbacks:\\

\begin{center}
{\it One challenge is to define the meaning, and measure the values, of inter-layer edges, and the choice of definition can itself play a significant role in the analyses... Furthermore,  intralayer  and  interlayer  edges  can represent ecological processes at different scales, and it is not always clear how to define the relative weight of interlayer edges with respect to intralayer edges...In ecology, this issue remains completely uncharted territory...\\
We also note that different types of interactions can also involve different `currencies’.  For example, pollination is measured differently than dispersal,  and it is important to consider discrepancies in the scales of the two edge types...}
\end{center}

Another alternative to model ecological systems is via agent-based models. In an agent-based model, a set of agents is defined, and a set of behavioral rules are defined for each type of agent in each of its possible states~\cite{Jannsen2006}. These models generally are applied to determine spatial dynamical patterns~\cite{Grimm2005}. In this sense, agent-based models are very interesting because it is possible to represent complex interaction mechanisms by means of a collection of behavioral rules, that once applied altogether (either sequentially or in parallel) represent the complex interaction, and one is also able to compute the dynamical evolution over space and time without much computational effort. However, there is no theoretical framework to study how structural properties of the rules in the model lead to stable configurations in the long-term, i.e. analytic methods to study the Complexity-Stability debate in agent-based models are poor. The usual strategy to gather knowledge about the stability of a system is to simulate the system under different configurations and then infer properties from the outputs of the simulation~\cite{Joop2010}. The problem with the latter strategy is that when we consider a system containing a large number of species and interactions, the number of parameters that one has to control becomes too large. Therefore, performing enough simulations to establish results about the stability of a large system is unfeasible in this approach.\\  
  
In table~\ref{MF1} we summarize the features of each of the modeling frameworks reviewed in this section.  

\begin{table}[h!]
\centering
\begin{tabular}{|c|c|c||c|c|c|}\hline
{\bf CS reps.} & Specs. & Interacts. & Dyn. Evo. & Mechanisms& Analytic Tools\\ \hline
Dyn. Eqs. & Few & Few & Direct & Yes & Rich\\ \hline
Networks & Many & Few & Indirect & No & Rich\\ \hline
Agent-based & Many & Many & Direct & Partial & Poor\\ \hline
\end{tabular}
\caption{Modeling languages applied to the Complexity-Stability debate. The first column specifies the modeling language, the second and third columns specify the feasible amount of species and interactions that the language is able to incorporate respectively. The fourth column specify if the language directly or indirectly incorporates dynamical evolution. The fifth row specifies if the language allows for a mechanistic description of the interactions and the sixth column specifies if the language is rich or poor in analytic tools.}
\label{MF1}
\end{table}

\section{Reaction Networks and the Modeling Ecological Interactions}

A reaction network is defined by a pair $(\Ms, \Rs)$, where $\Ms =\{a, b, c, \dots\}$ is a set of molecular species, and $\Rs \subseteq {\cal P}_m(\Ms)\times {\cal P}_m(\Ms)$ is their set of reactions, where ${\cal P}_m(\Ms)$ denotes the set of multisets of $\Ms$. For example, in the reaction network of figure~\ref{COT1}, reaction $r_1=a\to 2a$ represents a self-reproduction process of species $a$, reaction $r_2=a+c\to c$ represents the destruction of species $a$ out of the interaction of species $a$ and $c$, and $r_4=b+c\to b+2c$ represents the production of species $c$ catalized by $b$.

Note that, contrary to traditional network approaches which represent different ecological interactions by different types of links, reaction networks represent ecological interactions by specifying combinations of inputs that produce combinations of outputs. These inputs and outputs can be ecological species, resources, or also more abstract entities such as conditions for certain interactions to occur. Therefore, ecological interactions can be represented by means of interaction mechanisms represented by the reactions. This opens up the possibility to incorporate multiple entities and multiple types of interactions at once. In table~\ref{Eco-interactions} we represent examples of reactions representing the most common ecological interactions.

\begin{table}[h!]
\centering
\begin{tabular}{|c|c|}\hline
Reaction  & Ecological Interaction \\ \hline
$prey + predator \to 2 predator$ & Depredation\\ \hline
$host + hosted \to 2 hosted$ & Parasitism \\ \hline
$host + hosted \to host + 2 hosted$ & Comensalism \\ \hline
$host +  hosted \to host$ & Amensalism \\ \hline
${Coop}_1+{Coop}_2 \to 2{Coop}_1+ 2{Coop}_2$ & Mutualism  \\ \hline
$c_1+res\to 2c_1$; $c_2+res\to 2c_2$& Competition\\ \hline
\end{tabular}
\caption{Basic representation of ecological interactions in terms of reaction networks. In this simplified version, the first five ecological interactions are represented by a single reaction. The last row contains two reactions that representing competition between two species for a resource.}
\label{Eco-interactions}
\end{table}

The case of competition in table~\ref{Eco-interactions} is interesting because it illustrates in a very simple way that certain interaction mechanisms cannot be described by a single reaction. The mechanism describing an ecological interaction is in general represented by a reaction network. For example, we can provide a more detailed account of a possible mechanism underlying the mutualistic interaction in table~\ref{Eco-interactions}, assuming that the two species ${Coop}_1$ and ${Coop}_2$ create resources ${res}_1$ and ${res}_2$ respectively, that facilitate the other species' survival. This relation occurs for example between mychorrizae and plants~\cite{Harley1959}. In this case, mycorrhizae ${Coop}_1$ feeds from the roots ${res_2}$ of the plant ${Coop}_2$, and produces mycelium ${res}_1$, which in turn increments the absorption capacities of ${Coop}_2$. We can model this mutualistic relation by the set of reactions
\begin{equation}
 \begin{split}
{Coop}_2\to {Coop}_2+{res}_2&~\text{(Plant grow roots)} \\
{Coop}_1\to {Coop}_1+{res}_1&~\text{(Mychorrizea produces mycelium)}\\
{Coop_2}+{res}_1\to 2{Coop}_2&~\text{(Mycelium foster the growth of plants)}\\ 
{Coop_1}+{res}_2\to 2{Coop}_1&~\text{(Roots foster the growth of mychorrizea)}.
\label{MP-interaction}
\end{split}
\end{equation}

Note that other aspects such as the energy consumption, reproduction and death of mychorrizae, or pollination of plants are not specified in this simplified model. However, the reaction network model can be extended to not only provide such specification, but also to incorporate additional resources or species involved in finer grained descriptions of the interactions. Therefore, multiple ecological interactions can be specified with as much detail as needed, and integrated in a single reaction network representing the ecosystem. 

If we consider a realistic ecosystem model using reaction networks, we consider a reaction network with hundreds or thousands of species and interactions. Although from a reaction network it is possible to build a continuous, discrete and stochastic dynamical system that would allow to compute the evolution of the system, this dynamical approach is either computationally very expensive or have too many parameters to be studied analytically. In this sense the dynamical approach to study reaction networks falls in the same problems we encountered with differential equations and agent-based models in section~\ref{MEco}. Therefore, in order to apply reaction networks to the Complexity-Stability problem we need to find an alternative way to relate structure and dynamical stability.   
 	
\section{Chemical Organization Theory}
\label{COTM}
Chemical Organization Theory (COT)~\cite{Dittrich2007} is a biochemical inspired formalism whose aim is to study complex reaction networks. The interesting feature about COT is that it provides an elegant characterization of all the system's possible stable states. These states are put in correspondence with sets of species that hold particular properties, called organizations.

\begin{figure}[h!]
 \begin{center}  
  \includegraphics[height=4.5cm,width=9cm]{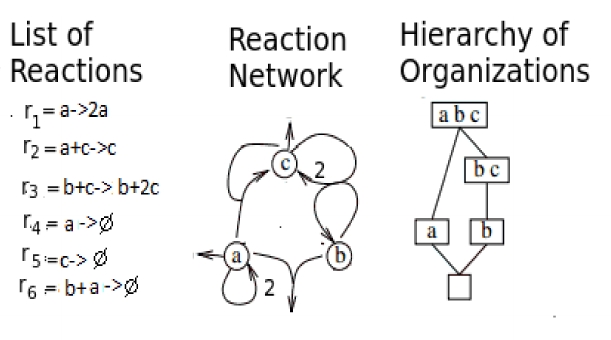}
\caption{Example of a reactions network, and its induced hierarchy of organizations.}
 \label{COT1}
 \end{center}
\end{figure}

An organization denotes a set of interacting species that are able to co-exist in the long term. This means that, although the system is constantly creating and destroying its own components, the complete set of species remains invariant because what disappears in one reaction is recreated by another, and no qualitatively new components are added. Interestingly, the set of organizations, called the organizational structure, forms a hierarchy where organizations at higher levels contain organizations of the lower levels (see figure~\ref{COT1}).

This notion of organization is close to the definition of autopoiesis, a concept that Maturana and Varela introduced to characterize living organisms~\cite{Razeto2012,Varela1974}. As such, organizations were introduced as a simple model for the origin of life out of interlocking cycles of chemical reactions~\cite{Benko2009}, and as a generalization of the well-studied notion autocatalytic set~\cite{Hordijk2010,Hordijk2018}. COT has been primarily used to analyze dynamical properties of chemical reaction networks, with a focus on the emergence of stable systems. The first examples were models of virus dynamics~\cite{Matsamaru2006}, and the chemistry of a planetary atmosphere~\cite{Centler2007}. A related application domain is the modeling of metabolic networks such as the bacterium E. Coli~\cite{Centler2007b}, and of genomic networks such as mitotic spindle assembly checkpoint~~\cite{Kreyssig2012}. Also various structural analysis connecting different approaches to metabolic, regulatory, and genomic networks have been developed~\cite{Contreras2011,Kaleta2006,Kaleta2008}.\\

Let  $(\Ms, \Rs)$ be a reaction network and consider a set of species $X\subseteq \Ms$. Note that some of the reactions in $\Rs$ may require species not included in $X$. Hence, the set $\Rs_X$ of reactions that can be triggered by the species in $X$ is in general smaller than $\Rs$. Hence, COT concerns with the sub-reaction networks $(X,\Rs_X)$ contained in the network $(\Ms,\Rs)$. In particular, the formal definition of an organization is derived from two characteristics about the stability of network $(X,\Rs_X)$:
\begin{enumerate}
 \item Closure: The resources produced by the reactions in $R_X$ are already in the starting set $X$. This means that no new molecules are added to $X$ by triggering the reactions in $\Rs_X$: 
\item Self-maintenance: All the reactions in $\Rs_X$ can operate at positive rates such that no reactant is consumed more than what is produced\footnote{We omit the mathematical formulation of this property for simplicity.}.   
 \end{enumerate}
The set $X$ is an organization if and only if is closed and self-maintaining. Therefore, organizations are dynamically invariant: no resources are added (closure) and no resources are removed (self-maintenance). 

For example, at the lower level of the hierarchy representing the organizational structure shown in Fig.~\ref{COT1}, we find that the sets $\{a\}$ and $\{b\}$ are organizations. While the former triggers $r_1$, and thus any reaction process\footnote{This is known in reaction network modeling as a flux vector. A flux vector is an specification of the relative rates of occurrence of reactions in the network.} where $r_1$ occurs at a larger or equal rate than $r_4$ makes it self-maintaining, the latter does not trigger any reaction, and hence it is trivially closed and self-maintaining (a non-reactive organization). At the next level of organizations, we only encounter the set $\{b,c\}$ which is self-maintaining when the rate of $r_3$ is larger or equal than the rate of $r_5$. 

From a technical perspective, it has been proven that the vast majority of stable regimes of a continuous dynamical system described by reaction networks, including fixed points~\cite{Dittrich2007}, and higher dimensional attractors such as periodic orbits and limit cycles~\cite{Peter2011}, correspond to organizations. Similar results can be obtained for the case of a discrete dynamical system~\cite{Kreissig2014}, and methods to rule out organizations that are dynamically unfeasible have also been developed~\cite{Veloz2011c}.

Therefore, COT characterizes the long-term dynamics of a reaction network as a process consisting of `movements between organizations' triggered by external perturbations. The perturbation, when understood as a slight change of state (in the dynamical systems sense), can cause that the stable regime leaves its basin of attraction, and thus the system will evolve towards another element of the organizational structure~\cite{Matsamaru2006}. Moreover, for structural perturbations, i.e. addition or elimination of novel species or interactions, COT also allows for an elegant representation of the dynamical change. When a structural perturbation occurs the organizational structure itself is modified. A decomposition theorem for organization allows to identify the parts of the organizational structure that are affected by the structural perturbation, and thus different algorithms have proved that computing the organizational structure is feasible for reactions networks of hundreds or even thousands of species~\cite{Centler2008,Veloz2017b,Veloz2019-BM}. 

Consider for example the network of figure~\ref{COT1} and the set $\{a,b,c\}$. If an external perturbation eliminates all the species of type $b$ from the system, we remain with the set $\{a,c\}$. However, this set is not an organization. Namely, $\{a,c\}$ is closed but not self-maintaining because $c$ is consumed by reaction $r_6$ but there is no way to recover species $c$ lost in $r_6$ through the activable reactions of $\{a,c\}$. Hence, the set of species $\{a,c\}$ will in the long-term evolve to the organization $\{a\}$ or to $\{\emptyset\}$ (depending on the rates of $r_1$ and $r_4$). Analogously, if we remove all the species of type $c$ from $\{a,b,c\}$, the system evolves in the long-term to the set $\{a\}$ or $\{\emptyset\}$. Finally, if we remove all the species of type $a$, the system remains at its perturbed configuration because $\{b,c\}$ is an organization (when the rate of $r_3$ is larger than the rate of $r_5$).\\



\section{COT and the Complexity-Stability Debate}
\label{COT-CS}
Reaction networks allow to develop mechanistic models of the interactions found in an ecosystem at any desired level of specification. These interactions can next be integrated in a reaction network model of the ecosystem. Since the reaction network modeling an ecosystem is going to be too large to be analyzed by dynamical systems' methods, we propose that COT provides a novel perspective to understand the dynamics of an ecological system. COT focuses on identifying collections of species (and resources) within our ``ecological universe'' whose structure allows them to function as a sustainable module (organization) with respect to the rest of the ecological universe. In this way one shifts the attention from ``what are the conditions that make a certain community to be stable?'' to ``given an ecological universe, what are the subsystems that form sustainable communities?''.
 
This change in perspective can be tremendously useful to study the Complexity-Stability when we think of large reaction networks that incorporate diverse types of interactions. In particular, since organizations at the higher levels of the hierarchy can be understood as combinations of organizations at lower levels, COT relates the recursive structure of organizations to the stability of the reaction network. For example, in~\cite{Veloz2017b} it is shown that organizations function in dynamically independent modules, and some of these modules are more fragile to perturbations than others, while in~\cite{Veloz2019-BM} it is found that the dynamical analysis of a certain organizations can be disregarded because their dynamical properties can be obtained from the dynamics of the smaller organizations that are contained in them. In this vein, complexity indicators such as synergy and non-decomposability have been developed to better understand those cases. Therefore, COT provides an interesting conceptual landscape to describe relation between complexity and stability by analyzing the inner structure of organizations and the properties of the organizational structure. In addition, there is a vast literature relating structural properties of reaction networks to their dynamical stability beyond COT. Other areas in the reaction network research such as flux-balance analysis~\cite{Orth2010}, elementary modes~\cite{Schuster1999}, metabolic cuts~\cite{Lacroix2008}, and several methods imported from other areas such as Petri-net research~\cite{Heiner2008} can also be incorporated to study the Complexity-Stability debate.
 
In table~\ref{MF2} we summarize the features of reaction networks as a language for modeling ecological interactions and studying the Complexity-Stability debate following the same schemata of table~\ref{MF1}

\begin{table}[h!]
\centering
\begin{tabular}{|c|c|c||c|c|c|}\hline
 {\bf CS reps.} & Specs. & Interacts. & Dyn.Evo & Mechanisms& Analytic Tools\\ \hline
 RN+COT& Many & Many & Indirect & Yes & Rich\\ \hline
\end{tabular}
\caption{Reaction Networks and COT applied to the Complexity-Stability debate.}
\label{MF2}
\end{table}

The COT approach to the Complexity-Stability debate takes the organizational structure, and the structure of organizations themselves, as the starting point to study the relation between complexity and stability. Hence, the organizational structure represents the collection of possible stable states to which the system can arrive in the long-term. These states are abstractions of the phase space, as there is no direct determination of the exact point or trajectory that the stable regime will occupy in the long-term dynamics. This abstract notion of stable state is indeed equivalent to the notion of subspace of a phase space~\cite{Dittrich2007}. Hence, COT identifies all the subspaces where stable regimes can be found, and equivalently, it discards all the subspaces of the phase space where stable regimes cannot be found.  
  
From this abstract notion of state, we can think in the sentence ``the system will evolve towards a certain organization'' as a proposition in the logical sense. In the next section we will show how this perspective shows an interesting connection between the Complexity-Stability debate and quantum theory.    

\section{COT and Non-Classical Ecological Structures} 
\label{QCOT}

In the early times of quantum theory, Birkhoff and Von Neumann realized that in a theory whose states are represented in a phase space, the subsets of the phase space play the role of propositions, and that in this sense, set inclusion corresponds to logical implication at the level of propositions. In their own words~\cite{VonNeumann1936}:\\ 

{\it ...in any physical theory involving a phase-space, the experimental propositions concerning a system $\Omega$ correspond to a family of subsets of its phase-space $\Sigma$, in such a way that ``$x$ implies $y$'' ($x$ and $y$  being any two experimental propositions) means that the subset of $\Sigma$ corresponding to $x$ is contained set-theoretically in the subset corresponding to $y$. This hypothesis clearly is important in proportion as relationships of implication exist between experimental propositions corresponding to subsets of different observation-spaces...\\
...Thus we see that the properties of logical implication are indistinguishable from those of set-inclusion, and that therefore it is algebraically reasonable to try to correlate physical qualities with subsets of phase-space...\\
...a system in which the relation ``$x$ implies $y$" is written $x \subset y$, is usually called a ``partially ordered system,'' and thus our first postulate concerning propositonal calculi is that the physical qualities attributable to any physical system form a partially ordered system.} \\ 

The latter paragraphs started the logico-algebraic approach to quantum theory, based in order-theoretical structures, principally in the theory of lattices~\cite{Garg2015}. The lattice approach to quantum theory has provided very important results and insights in the axiomatization of quantum theory, as well in its relation to epistemology and logic~\cite{Aerts2002,Mackey1963,Piron1976,VonNeumann1936}. 

In order to relate the latter ideas to COT, note that the organizational structure is a logical implication structure in this sense. Larger organizations contain smaller organizations, and thus the organizational structure can be seen as a partially ordered system. Moreover, it is possible to introduce certain operators known as `joint' and `meet' (which are a sort of generalization of the union and intersection), and the resulting structure is (for certain classes of reaction networks) a lattice~\cite{Dittrich2007,Speroni2015}. \\

Let $(\Ms,\Rs)$ be a reaction network, $\Os$ its the organizational structure, $X\subseteq \Ms$, and $G_{\Os}(X)$ the smallest organization containing $X$. The operator $G_\Os$ is called `generated organization', and for certain networks it can be proven that for each set $X$ its generated organization is unique~\cite{Dittrich2007}. For simplicity we will assume that $G_{\Os}(X)$ is unique for all $X$. Let the join $\js$ and meet $\ms$ operators be defined by $X\js Y=G_{\Os}(X\cup Y)$, and $X\ms Y=G_{\Os}(X\cap Y)$. Imposing a few extra axioms such that $X\js Y$ and $X\ms Y$ always exist and are unique, one can assert that $(\Os,\js,\ms)$ is a lattice. For simplicity, we will assume we are dealing with networks following those axioms, and refer to~\cite{Veloz2011b} for technical aspects concerning how to determine in which classes of networks its organizational structure forms a lattice. From here, it is is possible to establish properties about the structure of the lattice of organizations in the same way than Birkhoff and Von Neumann did for quantum systems, i.e. linking structure with logic. There is a huge amount of literature devoted to the axioms that a lattice has to hold in order to relate it to a particular logical structure~\cite{Mackey1963,Piron1976}. However, it is well known that distributivity is one of the crucial properties to discern whether or not the propositional structure obtained from a lattice represents a classical-logical structure. Namely, a lattice is distributive if and only if for any three elements $X,Y,Z$ in $\Os$ we have that
\begin{equation}
X\js(Y\ms Z)=(X\js Y)\ms (X\js Z)
\end{equation}

Non-distributive lattices correspond thus to propositional systems that do not conform with the rules of classical logic, and it has been proven that the truth valuations of such propositions cannot be represented by means of a classical probabilistic scheme~\cite{Beltrametti1995}. It is not the aim of this article to dig deeper into the relation between non-distributivity and classical or quantum logic. However, it is important to mention that non-distributivity is understood as a footprint of contextuality in quantum theory~\cite{Svozil2009}. Contextuality in quantum theory reflects the impossibility to obtain a coherent global description of the system, as the results obtained for one measurement can contradict global assumptions obtained about the system from other measurements. In our case, a measurement corresponds to identify whether a certain group of species is able to survive in the long-run (is an organization), and contextuality in this case can be intuitively understood because the long-term survival of the group of species under consideration is not independent to the other species in its environment, and thus the survival of the group of species is in many cases sensitive to its context. We will explain this idea with a simple example.   
 
\begin{figure}[h!]
 \begin{center}  
  \includegraphics[height=6cm,width=11cm]{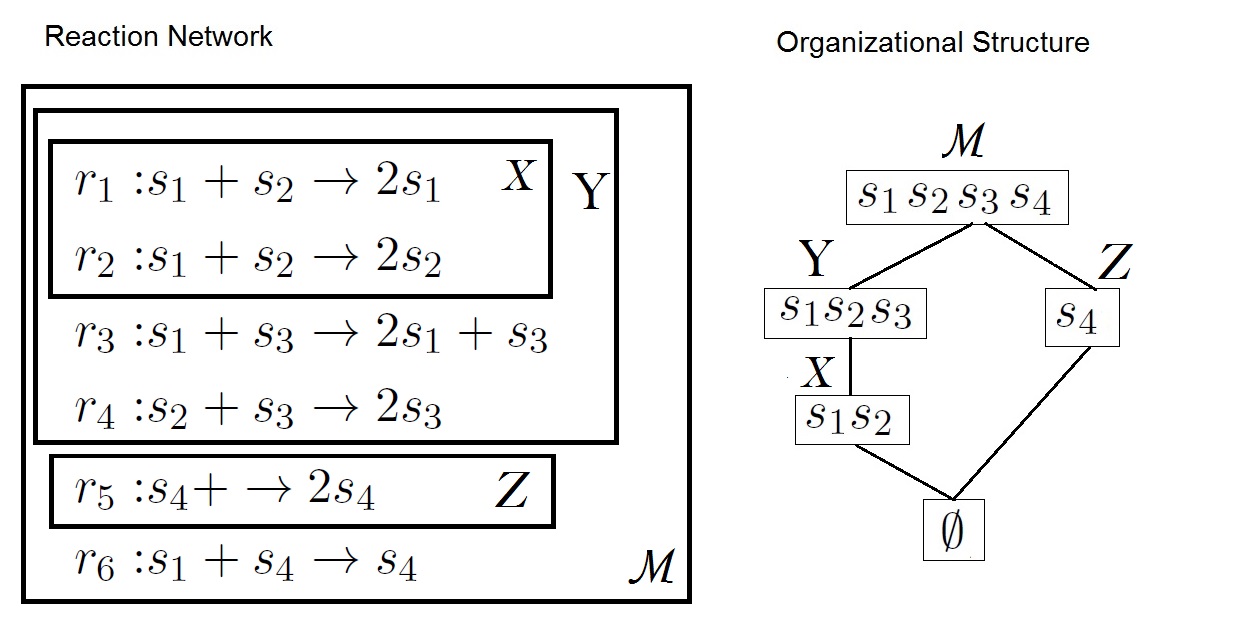}
\caption{Example of a reaction network, and its induced non-distributive organizational structure.}
 \label{Qeco}
 \end{center}
\end{figure}

In figure~\ref{Qeco} we show a reaction network model of an ecological system. Following the interactions described in table~\ref{Eco-interactions} we see that $s_1$ and $s_2$ can depredate each other, $s_3$ is in comensalistic relation with $s_1$ but depredates (or parasites) $s_2$, while $s_4$ is able to self-replicate and produces harm to $s_1$ without having any benefit (amensalism)\footnote{The ecological model we present in fig.~\ref{Qeco} might be a little unrealistic for an ecologist. However, it is important to note that this example aims at obtaining a non-distributive organizational structure for the smallest possible system. It is possible to find non-distributive organizational structures following only the interactions depicted in table~\ref{Eco-interactions}, but using a larger number of species and interactions.}

The organizational structure\footnote{Non-reactive organizations such as $\{s_1\}$ and $\{s_2\}$ are omitted for simplicity} in fig.~\ref{Qeco} is one of the prototypical non-distributive lattices, known as $N_5$. In fact, note that
\begin{equation}
\begin{split}
X\js(Y\ms Z)&=X\js G_{\Os}(Y\cap Z)=X\js\emptyset=X,\text{ and}\\
(X\js Y)\ms (X\js Z)&=G_{\Os}(X\cup Y)\ms G_{\Os}(X\cup Z)=Y\ms Z=\emptyset.
\end{split}
\end{equation}

The non-distributivity of this reaction network is explained by the fact that $s_1$, when considered as part of $X$, its growth is dependent on the growth of $s_2$. In fact, the self-maintainance of $X$ its conditioned to maximal production zero for both $s_1$ and $s_2$ (like a zero-sum game). However, when $s_1$ is considered as part of $Y$, its maximal production is in principle unbounded (due to $r_3$), and thus its growth is not limited by the growth of any other species. Then, $s_1$ cannot self-maintain when considered as part of $X$ interacting with $Z$ (due to reaction $r_6$), but when $s_1$ is considered as part of $Y$ interacting with $Z$ we have that $s_1$ can self-maintain. Indeed, $G_\Os(Y\js Z)=\Ms$. Therefore, the survival of $s_1$ is contextual to which group is considered to be part of. It is worth mentioning that we are not the first to identify contextuality in biological models~\cite{Aerts2010,Aerts2014,Melkikh2016,Real2016}. However, to our knowledge this is the first logico-algebraic relation between ecological systems and quantum structures based on reaction networks.   

\section*{Conclusion}
We introduced the language of reaction networks in ecology and proposed COT as an potential candidate to dilucidate the Complexity-Stability debate. Reaction networks 
represent situations among entities whose dynamical evolution can be understood as collective transformations of species. Intuitively, ecological systems are this type of situations. In particular, food webs represent the collective transformation of biomass among species, and other interactions can also be represented in a similar way (table~\ref{MF1}). Moreover, complex interaction mechanisms can be represented by means of reaction networks (see Eq.~\ref{MP-interaction} for an example). Since reaction network models of realistic ecosystems involve a large number of species and reactions, standard dynamical approaches in reaction networks are inadequate. For this reason, we propose COT as a potentially fruitful approach. In COT it is possible to obtain an elegant abstract and hierarchical dynamical landscape that accounts for the subsystems that are able to co-exist as independent modules (organizations). The internal structure of these subsystems can be studied from various mathematical perspectives, opening the possibility to apply results from computational biochemistry, systems biology, and others, thus enriching the toolkit with which the relation between structure and stability has been studied up to now. In particular, we presented an example of how the organizational structure can be seen as a logico-algebraic structure. The latter example opens up the possibility to explore the Complexity-Stability problem applying techniques from quantum theory that have been unprecedently applied in ecology. 

For future work, several lines of research can be developed. First, it is important to explore the modeling of complex ecological interactions by means of mechanisms and how the data collected by ecologists can be translated to reaction networks. It is interesting that the interaction mechanisms are in the end {\it narratives} about how species and resources interact. Therefore, ecologists need to work together to provide unified views on the mechanisms explaining how species interact. An attempt in this vein is a model of endosymbiotic interactions incorporating different representation layers~\cite{Veloz2019F1}. Second, it is necessary to develop measures of stability indicators such as resilience, robustness, adaptivity and so in the language of reaction networks. In~\cite{Veloz2017b} and~\cite{Heylighen2015} some ideas have been advanced in this respect. Third, it is important to extend COT to represent ecological interactions in space. This can be done extending the reaction network formalism to incorporate compartments~\cite{Ferellman2014,Veloz2011c}. Fourth, it is extremely important to build a common data-framework to represent mechanistic ecological interactions. We believe that one can leverage from notational schemes such as SBML for this purpose~\cite{Hucka2003}. In this sense, we could have in the future a database of reaction networks representing the different ecological interactions mechanisms that occur among every possible group of species, so we can build and analyze models integrating the relevant species and interactions of our purpose (similar to how metabolic reaction networks are analyzed today).

In summary, we believe that the application of reaction networks and COT to study large ecosystems has the potential to become an extremely important method to advance in the understanding between complexity and stability in ecology, and might even revolutionize the methodologies and understanding of general complex systems.\\

{\bf Acknowledgments.} This work was supported by the postdoctoral project Fondecyt 3170122.


\end{document}